\documentstyle[12pt]{article}

\makeatletter
\def\section{\@startsection{section}{1}{\z@}{-3.5ex plus -1ex minus -.2ex}
{2.3ex plus .2ex}{\large\bf}}
\makeatother
 
\makeatletter
\@addtoreset{equation}{section}

\makeatother

\def\la{\langle}
\def\ra{\rangle}

\def\lb{\lbrack}
\def\rb{\rbrack}

 \def\Slash#1{
  \begin{picture}(5,6)(0,0)
  \put(-.7,-1.2){\line(5,6)6}
  \end{picture}
  \kern-.8em#1}
 \def\slash#1{
  \begin{picture}(5,6)(0,0)
  \put(-1.5,-1.7){\line(5,6)5}
  \end{picture}
  \kern-.8em#1}

\def\Sn{\Slash \nabla}
\def\sd{\Slash \partial}

\def\Tr{\mbox{Tr}}
\def\tr{\mbox{tr}}
\def\e{\epsilon}
\def\g5{\gamma_5}
\def\hg5{\hat{\gamma}_5}
\def\A{{\cal A}}
\def\index{\mbox{index}\,}

\def\F{{\cal F}}
\def\C{{\cal C}}

\def\G{{\cal G}}
\def\U{{\cal U}}
\def\E{{\cal E}}
\def\hC{\widehat{\cal C}}
\def\wA{\widetilde{\cal A}}
\def\wF{\widetilde{\cal F}}

\def\ch{\mbox{ch}}
\def\wB{\widetilde{B}}
\def\wE{\widetilde{E}}
\def\wnabla{\widetilde{\nabla}}

\def\Qlatmr1{Q_{lat}^{(m=r=1)}}

\def\be{\begin{eqnarray}}
\def\ee{\end{eqnarray}}

\parskip=\medskipamount

\def\qed{\vrule height 1.5ex width 1.2ex depth -.1ex}

\begin{document}

\vspace{8mm}

\begin{center}
 
{\Large \bf Families index theory for Overlap lattice Dirac operator. I.}
\\

\vspace{0.3ex}

\vspace{12mm}

{\large David H. Adams}

\vspace{4mm}

Physics dept., National Taiwan University, Taipei, Taiwan 106, R.O.C.\\

and \\

Math. dept. and Centre for the Subatomic Structure of Matter, \\
University of Adelaide, S.A. 5005, Australia. \\

\vspace{1ex}

email: dadams@maths.adelaide.edu.au

\end{center}

\begin{abstract}

The index bundle of the Overlap lattice Dirac operator over the orbit 
space of lattice gauge fields is introduced and studied.
Obstructions to the vanishing of gauge anomalies in the Overlap formulation
of lattice chiral gauge theory have a natural description in this context.
Our main result is a formula for the topological charge (integrated Chern 
character) of the index bundle over even-dimensional spheres in the orbit 
space. It reduces under suitable conditions
to the topological charge of the usual (continuum)
index bundle in the classical continuum limit (this is announced and sketched
here; the details will be given in a forthcoming paper).
Thus we see that topology of the index bundle of the Dirac operator over 
the gauge field orbit space can be captured in a finite-dimensional lattice 
setting.

\end{abstract}

\section{Introduction}

The index bundle of the Dirac operator over the orbit space of gauge fields 
was studied many years ago by Atiyah and Singer \cite{AS}.\footnote{
Detailed treatments of Dirac operator index theory can be found in 
\cite{reviews}.} Obstructions
to the vanishing of chiral gauge anomalies have a natural description in
this context \cite{AS,AG}. 
More recently, the overlap formalism \cite{ov}, developed by Narayanan and
Neuberger as a way to formulate chiral gauge theories on the lattice, 
has provided the ingredients for
a lattice version of Dirac operator index theory when the base manifold
(spacetime) is an even-dimensional torus. A lattice version of the
index arose there as the fermionic topological charge of the lattice
gauge field, and can be expressed as the index of the
Overlap lattice Dirac operator, introduced in \cite{Neu1}. The fact that the
overlap formulation successfully reproduces the
global gauge anomaly and obstructions to the vanishing of local gauge 
anomalies \cite{Neu(SU(2)),Neu(geom),BC,Bar,DA(NPB)} indicates that
it should also lead to a lattice version of {\em families} index theory for 
the Dirac operator. The purpose of the present paper is to show that this is 
indeed the case.\footnote{A first step in this direction was made earlier
in \cite{DA(PRL)}.}

We develop a families index theory for the Overlap lattice Dirac operator
coupled to SU(N) lattice gauge fields on the 4-torus. 
(The specialisation to 4 dimensions is for simplicity and physical relevance;
everything generalises straightforwardly to arbitrary even-dimensional torus
\cite{prep}.)
After recalling the lattice setup in \S2, we show in \S3 that an index bundle 
over the orbit space of lattice gauge fields can be defined in a natural way.
In \S4 we derive our main result:
a formula (``lattice families index theorem'') for the 
topological charge (integrated Chern character) 
of this index bundle over a generic 
even-dimensional sphere in the orbit space of lattice gauge fields (Theorem 1,
eq. (\ref{4.1}) below). It contains an integer-valued part which has 
no continuum analogue; in \S5 we outline the argument in the forthcoming 
paper \cite{prep} showing that the 
rest of the expression reduces to the topological charge of the usual 
(continuum) index bundle in the classical continuum limit (Theorem 2).

When the group $\G$ of gauge transformations is restricted to 
$\G_0=\{\phi\in\G\;|\;\phi(x_0)\!=\!1\}$ (where $x_0$ is some basepoint in
the 4-torus) the orbit space $\U/\G_0$ is a smooth manifold (since $\G_0$ 
acts freely on the space $\U$ of lattice gauge fields). In this case
the above-mentioned part of the families index formula which has no continuum
analogue vanishes in the classical continuum limit, as we will see in \S5.
The above-mentioned classical continuum limit result
then implies that the finite-dimensional lattice setting is capturing topology 
of the continuum index bundle over $\A/\G_0$ when the lattice is sufficiently 
fine.

In \S6 we discuss how obstructions to the vanishing of gauge 
anomalies in the overlap formulation of lattice chiral gauge theories have 
natural descriptions in this context. 
We conclude in \S7 by discussing some implications and possible
generalisations of our results, and some related results to be shown in
subsequent papers. The background material in \S2-3 is intended to make this 
paper accessible to nonspecialists in lattice gauge theory.

This is the first in a series of papers. In the follow-up paper \cite{prep},
we treat the general case of arbitrary even-dimensional torus and prove the
classical continuum limit result announced and outlined in \S5, as well as
a result mentioned in the concluding remark 2.
In \cite{DA(app)}, we apply the results of the present paper
to demonstrate an interplay between topological features of 
the space of SU(N) lattice gauge fields on $T^4$ and the
existence question for $\G_0$ gauge fixings on the lattice which do not have 
the Gribov problem. We find that certain obstructions to the existence of
such gauge fixings in the continuum are absent on the lattice, and that 
instead the topological sectors
(specified by the fermionic topological charge) contain noncontractible
spheres of various even dimensions when $N\ge3$, and noncontractible
circles in the $N\!=\!2$ case. Further applications of the lattice families 
index theory developed here to the global gauge 
anomaly on the lattice and related issues will be given in \cite{DA(global)}.

\section{Lattice setup}

We take the spacetime manifold to be the Euclidean 4-torus $T^4$ and the
gauge group to be SU(N). A hypercubic lattice on $T^4$ determines the
space $\C$ of lattice spinor fields, the space $\U$ of lattice gauge fields,
and the group $\G$ of lattice gauge transformations.
$\C$ is the finite-dimensional complex vector space consisting of the
spinor-valued functions on the lattice sites, i.e. the functions on the
lattice sites with values in ${\bf C}^4\otimes{\bf C}^N$. (${\bf C}^4$ and
${\bf C}^N$ are the spin and SU(N) representation spaces; for simplicity
we are specialising to the fundamental representation of SU(N).)
It has the inner product
\be
\la\psi_1\,,\psi_2\ra=a^4\sum_x\psi_1(x)^*\psi_2(x)
\label{1}
\ee
($a$=lattice spacing). $\U$ consists of the SU(N)-valued functions on the
links of the lattice. It has the finite-dimensional manifold structure
\be
\U\;\cong\;\mbox{SU(N)}\times\mbox{SU(N)}\times\cdots\times\mbox{SU(N)}
\label{2}
\ee
(one copy for each link). The group $\G$ consists of the maps
$\phi:\{\mbox{lattice sites}\}\to\mbox{SU(N)}$. It acts on $\C$ and $\U$ by
\be 
\phi\cdot\psi(x)&=&\phi(x)\psi(x) \label{3} \\
(\phi\cdot{}U)_{\mu}(x)&=&\phi(x)U_{\mu}(x)\phi(x+ae_{\mu})^{-1}
\label{4}
\ee
($e_{\mu}$=unit vector in the positive $\mu$-direction).
Continuum spinor fields, gauge fields and gauge transformations have 
lattice transcripts, defined in a natural way: In the case of spinor field
$\psi(x)$ or gauge transformation $\phi(x)$ we restrict $x$ to the lattice 
sites to get elements in $\C$ or $\G$, respectively. 
In the case of gauge field
$A_{\mu}(x)$ the lattice transcript $U_{\mu}(x)$ is the parallel transport
from $x+ae_{\mu}$ to $x$. Then (\ref{3})--(\ref{4}) are the lattice 
transcripts of the continuum gauge transformations. 
For a given continuum gauge field $A$, the {\em classical continuum limit}
of a quantity depending on the lattice gauge field $U$ (e.g., the index of
a lattice Dirac operator) is the limit of infinitely many
subdivisions of the hyper-cubic lattice (i.e. lattice spacing $a\to0$)
with $U$ being the lattice transcript of $A$. The classical continuum limit
of quantities depending on families of lattice gauge fields is defined 
analogously.

The covariant derivative $\partial_{\mu}^A=\partial_{\mu}+A_{\mu}$
can be approximated on the lattice by the covariant finite difference 
operators $\frac{1}{a}\nabla_{\mu}^{\pm}\,$, $\frac{1}{a}\nabla_{\mu}$
where
\be
\nabla_{\mu}^+\psi(x)&=&U_{\mu}(x)\psi(x+ae_{\mu})-\psi(x) \label{5} \\
\nabla_{\mu}^-\psi(x)&=&\psi(x)-U_{\mu}(x-ae_{\mu})^{-1}\psi(x-ae_{\mu}) 
\label{6}
\ee
and $\nabla_{\mu}=\frac{1}{2}(\nabla_{\mu}^++\nabla_{\mu}^-)$. Note that
$(\nabla_{\mu}^{\pm})^*=-\nabla_{\mu}^{\mp}\,$, 
so $\nabla_{\mu}^*=-\nabla_{\mu}$.

\section{Lattice version of Dirac operator index theory}

The Dirac operator $\sd^A=\gamma^{\mu}\partial_{\mu}^A$ has the naive lattice
approximation
\be
\Sn=\gamma^{\mu}{\textstyle \frac{1}{a}}\nabla_{\mu}
\label{7}
\ee
However, the index theory for this operator is trivial: $\index\,\Sn^U=0\;$
$\forall\,U\in\U$. This well-known fact is a simple consequence of the
chiral symmetry
\be 
\g5\Sn+\Sn\g5=0
\label{8}
\ee
and the finite-dimensionality of $\C$. A nontrivial lattice version of the
Dirac operator index can instead be constructed by noting that 
$\index\,\sd^A$ is minus the spectral flow of $H_m^A=\g5(\sd^A-m)$
as $m$ increases from $m<0$ to $m>0$. (We are following the physics 
convention where the $\gamma^{\mu}$'s are hermitian; then $\sd^A$ is
antihermitian and $H_m^A$ is hermitian.) A lattice analogue of $H_m^A$ is
\be
H_m^U=\g5(D_w^U-m)
\label{9}
\ee
where
\be
D_w^U=\Sn^U+a{\textstyle \frac{r}{2}}\Delta^U
\label{10}
\ee
is the Wilson--Dirac operator \cite{Wilson}. The Wilson term 
$a\frac{r}{2}\Delta$, where 
$\Delta=\frac{1}{a^2}(\nabla_{\mu}^{\pm})^*\nabla_{\mu}^{\pm}=
\frac{1}{a^2}\sum_{\mu}\nabla_{\mu}^--\nabla_{\mu}^+$ is the lattice Laplace
operator and $r>0$ the Wilson parameter, breaks the chiral symmetry and 
thereby allows for a nontrivial index theory. It is known that when $m<0$
the hermitian operator $H_m^U$ has symmetric spectrum and no zero-modes
\cite{ov}. Hence the spectral flow as $m$ increases from any $m_1<0$ to
some $m_2>0$ is equal to half the spectral asymmetry of $H_m^U$ at 
$m\!=\!m_2\,$, i.e. $\frac{1}{2}\Tr(\e^U)\,$, where
\be
\e^U=\frac{H^U}{|H^U|}
\label{11}
\ee
This suggests defining the lattice version of the index by
\be
\index\,\sd^A\;\to\;-\frac{1}{2}\Tr(\e^U)
\label{12}
\ee
for some suitable value of the parameter $m$ in (\ref{9}) (which we have
suppressed in the notation in (\ref{11})--(\ref{12})). 
In fact the right-hand side of (\ref{12}), with 
\be
m=\frac{rm_0}{a}\qquad,\qquad\quad{}0<m_0<2
\label{13}
\ee
is precisely the definition of the topological charge of
$U$ which arose in the overlap formulation of chiral gauge theories on the
lattice \cite{ov}. In \cite{DA(JMP)} this was shown to reduce to 
$\index\,\sd^A$ in the classical continuum limit for any choice of $m$
satisfying (\ref{13}) above. It is also these choices of $m$ which give 
the correct classical continuum limit for the families index theory that
we develop in the following.

For this definition of the lattice version of the index we must exclude
from $\U$ the lattice gauge fields $U$ for which $H^U$ has zero-modes,
so that $\e^U$ is well-defined. This determines a decomposition of $\U$
into topological sectors labelled by the lattice index. It has been shown
in \cite{local} (see also \cite{Neu(bound)}) that $H^U$ has no zero-modes
when each $U(p)$ is sufficiently close to the identity, where $U(p)$ is
the product of the $U_{\mu}(x)$'s around a plaquette $p$ in the lattice.
(We refer to \cite{Neu(bound)} for the currently best bound on 
$||1-U(p)||$ which ensures this.) Consequently, the absence of zero-modes
for $H^U$ is guaranteed close to the classical continuum limit, since
when $U$ is the lattice transcript of a continuum field $A$, and $p$ is
the plaquette specified by a lattice site $x$ and directions $\mu$ and 
$\nu\,$, we have $1-U(p)=a^2F_{\mu\nu}(x)+O(a^3)$. (See \cite{DA(JMP)} for 
a more detailed discussion of this point.) From now on, we take $m$ as in 
(\ref{13}) and assume that a choice of $m_0\in(0,2)$ has been made, and 
$\U$ denotes the space of lattice gauge fields with the $U$'s for which 
$H^U$ has zero-modes excluded.

The lattice version (\ref{12}) of $\index\,\sd^A$ can be expressed as the
index of the Overlap lattice Dirac operator \cite{Neu1}, given by
\be
D={\textstyle \frac{1}{a}}(1+\g5\e)
\label{14}
\ee
with $\e$ as in (\ref{11}). The nullspace of this operator is invariant under
$\g5$ and therefore has a chiral decomposition
\be
\ker{}D^U=(\ker{}D^U)_+\oplus(\ker{}D^U)_-
\label{16}
\ee
The index can then be defined as 
\be
\index\,D^U=\dim(\ker{}D^U)_+-\dim(\ker{}D^U)_-
\label{17}
\ee
and coincides with the above lattice version of $\index\sd^A\,$:
\be
\index\,D^U=-\frac{1}{2}\Tr(\e^U)
\label{18a}
\ee
While this was obtained directly in \cite{Neu1}, it can also be seen using 
the fact that (\ref{14}) satisfies \cite{Neu2} 
\be
\g5{}D+D\g5&=&aD\g5{}D\qquad\qquad\mbox{(GW relation)} \label{15a} \\
D^*&=&\g5{}D\g5\qquad\qquad\mbox{($\g5$-hermiticity)} \label{15b}
\ee
The first relation was originally studied by
Ginsparg and Wilson \cite{GW}, and later rediscovered by Hasenfratz
and collaborators \cite{laliena}, who noted that the nullspace of a
solution $D$ is invariant under $\g5$ (since if $D\psi=0$
then $D\g5\psi=(aD\g5{}D-\g5{}D)\psi=0$). Furthermore, they showed that
solutions of (\ref{15a})--(\ref{15b}) satisfy the following index 
formula (see also \cite{L(PLB)}):
\be
\index\,D^U=-\frac{a}{2}\Tr(\g5{}D^U)
\label{18b}
\ee
Substituting (\ref{14}) in (\ref{18b}) leads to (\ref{18a}).

Having reviewed the previously developed lattice index theory for the Dirac
operator we now proceed to develop a lattice version of the {\em families}
index theory. Rather than the index (\ref{17}), the central object here
is the index {\em bundle}, formally given, in analogy with the continuum
case, by
\be
\mbox{``$\ \index\,D=(\ker{}D)_+-(\ker{}D)_-\ $''}
\label{19}
\ee
where
\be
(\ker{}D)_{\pm}=\{(\ker{}D^U)_{\pm}\}_{U\in\U}
\label{20}
\ee
If $(\ker{}D)_+$ and $(\ker{}D)_-$ were vector bundles  then (\ref{19})
would be a well-defined element in $K(\U)$, the K-theory of $\U$.
However, as in the continuum setting, the dimensions of 
$(\ker{}D^U)_+$ and $(\ker{}D^U)_-$ can jump as $U$ varies, even though their
difference, $\index\,D^U$, remains constant. In the continuum some
trickery is required to deal with this aspect so as to make the index bundle
into a well-defined element in the K-theory (see, e.g., \cite{reviews}).
But in the lattice setting
this aspect can be dealt with in a simple way, by exploiting the 
finite-dimensionality of $\C$, as follows.
Besides the usual chiral decomposition,
\be
\C=\C_+\oplus\C_-\qquad\qquad(\mbox{$\g5=\pm1$ on $\C_{\pm}$})\,,
\label{21}
\ee
there is another decomposition determined by $\e^U$ (which played a central
role in the overlap formalism \cite{ov}, see also 
\cite{L(abelian),L(nonabelian)}):
\be
\C=\hC_+^U\oplus\hC_-^U\qquad\qquad(\mbox{$-\e^U=\pm1$ on $\hC_{\pm}^U$})
\label{22}
\ee
By (\ref{14}),
\be
-\e^U=\g5(1-aD^U)
\label{23}
\ee
which looks formally like a gauge field-dependent lattice deformation of
$\g5$, hence the ${\pm}$ convention in (\ref{22}).
It follows from (\ref{23}) that
\be
(\ker{}D^U)_{\pm}\;\subset\;\hC_{\pm}^U\cap\C_{\pm}
\label{24}
\ee
In light of this we can define $V^U$ to be the orthogonal complement of 
$(\ker{}D^U)_+$ in $\hC_+^U\,$, and $W^U$ the orthogonal complement of 
$(\ker{}D^U)_-$ in $\C_-\,$, i.e.
\be
\hC_+^U&=&(\ker{}D^U)_+\oplus{}V^U \label{25} \\
\C_-&=&(\ker{}D^U)_-\oplus{}W^U\,.
\label{26}
\ee
From (\ref{14}) (or (\ref{15a}) and (\ref{23})) we get
\be
\g5{}D^U=D^U\e^U
\label{27}
\ee
which implies that $D^U$ maps $\hC_{\pm}^U$ to $\C_{\mp}$.

\vspace{1ex}
\noindent {\em Lemma 1}. The map $D^U:\hC_+^U\to\C_-$ restricts to an
isomorphism $D^U:V^U\stackrel{\cong}{\to}W^U$.
\vspace{1ex}

\noindent {\em Proof}. It suffices to show that the orthogonal complement
of the image of $D^U$ on $\C$ is precisely $\ker{}D^U$. But this is a 
simple consequence of the $\g5$-hermiticity property (\ref{15b}) and
the invariance of $\ker{}D^U$ under $\g5$. \qed

\noindent Now set
\be
V=\{V^U\}_{U\in\U}\quad\quad,\quad\quad{}W=\{W^U\}_{U\in\U}
\label{28}
\ee
These are not vector bundles in general, since $\dim{}V^U$ and $\dim{}W^U$
can jump as $U$ varies, but we will formally treat them as vector bundles
in the following. By Lemma 1, $V$ and $W$ are isomorphic. Using this and
(\ref{25})--(\ref{26}), formal K-theoretic manipulations give
\be
\index\,D&=&(\ker{}D)_+-(\ker{}D)_- \nonumber \\
&\cong&(\ker{}D)_+\oplus{}V-(\ker{}D)_-\oplus{}V \nonumber \\
&\cong&(\ker{}D)_+\oplus{}V-(\ker{}D)_-\oplus{}W \nonumber \\
&=&\hC_+-\C_-
\label{29}
\ee
where 
\be
\hC_+=\{\hC_+^U\}_{U\in\U}
\label{30}
\ee
and $\C_-$ in (\ref{29}) is to be interpreted as the trivial vector bundle 
over $\U$ with fibre $\C_-$.
Unlike the initial expression, the final expression (\ref{29}) for
$\index{}D$ is a well-defined element in $K(\U)$ due to
the following:

\vspace{1ex}
\noindent {\em Lemma 2}. $\hC_+$ is a smooth vector bundle over $\U$.
\vspace{1ex}

\noindent {\em Proof}. Since $\dim{}V^U=\dim{}W^U$ we have 
\be
\index\,D^U=\dim\hC_+^U-\dim\C_-
\label{31}
\ee
hence $\dim\hC_+^U$ is locally constant under variations of $U$.
Furthermore, $\hC_+={\cal P}(\C)$ where $\C$ is to be interpreted as 
the trivial vector bundle over $\U$ with fibre $\C$ and 
${\cal P}:\C\to\C$ is the smooth 
vector bundle map defined on the fibres by
\be
P^U={\textstyle \frac{1}{2}}(1-\e^U)
\label{32}
\ee
Then by a standard mathematical result (see, e.g., Prop. 1.3.2.(ii)
of \cite{Atiyah(K)}), $\hC_+$ is a smooth vector bundle over $\U$. \qed

We therefore take (\ref{29}) as the definition of the index bundle of $D$.
Due to gauge covariance it descends to an element in the K-theory of the
orbit space, i.e.
\be
\index\,D\;\in\;K(\U/\G)
\label{33}
\ee
Due to the triviality of $\C_-$ in (\ref{29}), the topology of the index 
bundle is determined solely by $\hC_+$. In particular, for the 
nonzero degree parts of the Chern
character we have $\ch_n(\C_-)=0$ and\footnote{The zero degree part of 
the Chern character
is just $\ch_0(\index{}D)=\dim\hC_+^U-\dim\C_-=\index{}D^U$.}
\be
\ch_n(\index\,D)=\ch_n(\hC_+)-\ch_n(\C_-)=\ch_n(\hC_+)\quad\qquad(n\ge1)
\label{34}
\ee
Thus the topological charge of the index bundle
over general $2n$-dimensional spheres in the orbit space, which is obtained by
integrating the Chern character, is the same as for the vector bundle
$\hC_+$ for $n\ge1$.
We remark that, since $\hC_+$ is a unitary vector bundle
(with unitary structure determined by the inner product (\ref{1})),
its topological charge over odd-dimensional spheres always vanishes.
So it is only the even-dimensional case that is of interest.

\section{Topological charge of the index bundle over 2n-spheres in the
orbit space}

Generically, a $2n$-sphere in $\U/\G$ can be presented as a $2n$-dimensional
family $U^{(\theta,t)}$ of lattice gauge fields, parameterised by the
$2n$-ball $B^{2n}$ with boundary $S^{2n-1}\,$ ($\theta\in{}S^{2n-1}$ and
$t$ is the radial coordinate in $B^{2n}$), with all the boundary points
$U^{(\theta,1)}$ being gauge equivalent, i.e. 
$U^{(\theta,1)}=\phi_{\theta}\cdot{}U\,$ ($U\equiv{}U^{(0,1)}$) for some
family $\{\phi_{\theta}\}_{\theta\in{}S^{2n-1}}$ of lattice gauge 
transformations. Set $P^{(\theta,t)}=P^{U^{(\theta,t)}}$
(given by (\ref{32})). The main result of this paper is the following 
formula, which can be regarded as a ``lattice families index theorem'':

\vspace{1ex}
\noindent {\em Theorem 1}. The topological charge of $\hC_+\,$ -- and 
therefore of the index bundle -- over the above $2n$-sphere in $\U/\G$ 
($n\ge1$) is
\be
Q_{2n}&=&{\textstyle \frac{1}{(2\pi{}i)^n}}
\left({\textstyle \frac{1}{n!}}\int_{B^{2n}}
\Tr\left\lb{}P^{(\theta,t)}(dP^{(\theta,t)})^{2n}\right\rb 
\,-\,{\textstyle \frac{(-1)^n}{2}\frac{(n-1)!}{(2n-1)!}}
\int_{S^{2n-1}}\Tr\left\lb\e^U
(\phi_{\theta}^{-1}d_{\theta}\phi_{\theta})^{2n-1}\right\rb
\ \right) \nonumber \\
&&\;+\ 2\sum_xdeg(\phi(x))
\label{4.1}
\ee
where $\Tr$ is the trace for linear operators on $\C\,$,
$\,d\!=\!d_{\theta}\!+\!d_t$ is the exterior derivative on $B^{2n}$ and
$deg(\phi(x))$ is the degree of the map $S^{2n-1}\to\mbox{SU(N)}\,$,
$\,\theta\mapsto\phi_{\theta}(x)$.

\vspace{1ex}

\noindent {\em Remark}.
The last term $2\sum_xdeg(\phi(x))$ vanishes in the $n\!=\!1$
case (since SU(N) contains no noncontractible circles), but can be 
nonvanishing for $2\le{}n\le{}N$. This even integer-valued term has no
continuum analogue. The rest of (\ref{4.1}) reduces to the
topological charge of the continuum Dirac index bundle in the classical
continuum limit, cf. \S5. We will also see there that when
$\G$ is replaced by the constrained group of gauge transformations $\G_0\,$, 
for which the orbit space is a smooth manifold, the term $2\sum_xdeg(\phi(x))$
gives no contribution to the classical continuum limit.

\vspace{1ex}
\noindent {\em Proof of Theorem 1}. 
The restriction of the vector bundle $\hC_+$ to the 
$2n$-sphere in $\U/\G$ has the following characterisation. 
Set $\hC_+^{(\theta,t)}:=\hC_+^{U^{(\theta,t)}}$ and define the vector bundle
\be
E:=\{\hC_+^{(\theta,t)}\}_{(\theta,t)\in{}B^{2n}}
\label{34a}
\ee
over $B^{2n}$. The action of the gauge transformation
$\phi_{\theta}$ on $\C$ restricts to an isomorphism
\be
\phi_{\theta}:\hC_+^U\stackrel{\cong}{\to}\hC_+^{(\theta,1)}
\label{35}
\ee
This determines an equivalence relation $\sim$ identifying each 
$\hC_+^{(\theta,1)}$ with $\hC_+^U$. The bundle $\hC_+$ over the $2n$-sphere
in $\U/\G$ is then given by $E/_{\sim}$. 
Topologically, this is equivalent to
the bundle over $S^{2n}$ constructed as follows. Let $\wB^{2n}$ denote
another copy of the $2n$-ball, with coordinates $(\theta,s)$, and let
$\wE$ denote the trivial bundle over $\wB^{2n}$ with fibre $\hC_+^U$.
Then $E$ and $\wE$ can be glued together along the common boundary $S^{2n-1}$
of $B^{2n}$ and $\wB^{2n}$ via (\ref{35}) to get a bundle $\E$ over
$S^{2n}=B^{2n}\cup_{S^{2n-1}}\wB^{2n}$ with the same topology as the 
restriction of $\hC_+$ to the $2n$-sphere in $\U/\G$.

To get a formula for the topological charge we introduce a covariant 
derivative (connection) in $\E$. 
With $d=d_{\theta}+d_t\,$,  
\be
\nabla^{(\theta,t)}=P^{(\theta,t)}\circ{}d\circ{}P^{(\theta,t)}
\label{36}
\ee
is a connection in the bundle $E$ over $B^{2n}\,$,
and, with $d=d_{\theta}+d_s\,$, 
\be
\wnabla^{(\theta,s)}=P^U\circ(
d+f(s)\phi_{\theta}^{-1}d_{\theta}\phi_{\theta})\circ{}P^U
\label{37}
\ee
is a connection in the bundle $\wE$ over $\wB^{2n}\,$,
where $f$ is some cutoff function with $f(s)\!=\!1$ and $f(s)\!=\!0$
in neighborhoods of $s\!=\!1$ and $s\!=\!0$, respectively.
A simple calculation shows that
\be
\nabla^{(\theta,1)}=\phi_{\theta}\circ
\wnabla^{(\theta,1)}\circ\phi_{\theta}^{-1}
\label{38}
\ee
Consequently, $\nabla$ and $\wnabla$ fit together to give a connection
in $\E$. Hence the topological charge of this bundle is
\be
Q_{2n}=\frac{1}{(2\pi{}i)^n\,n!}\,\left\lb\ \int_{B^{2n}}\Tr\,(F_{\nabla})^n
-\int_{\wB^{2n}}\Tr\,(F_{\wnabla})^n\ \right\rb
\label{39}
\ee
where $F_{\nabla}$ and $F_{\wnabla}$ are the curvatures of $\nabla$ and
$\wnabla\,$, respectively.

Using the fact that for general connection $\nabla$ we have
$\nabla\circ\nabla=F_{\nabla}$ (i.e. wedge multiplication by $F_{\nabla}$),
simple calculations in the present case give
\be
F_{\nabla}^{(\theta,t)}&=&P^{(\theta,t)}dP^{(\theta,t)}dP^{(\theta,t)}
\label{40} \\
(F_{\nabla})^n&=&P(dP)^{2n}
\label{41}
\ee
and
\be
F_{\wnabla}^{(\theta,s)}
&=&\left\lb\,f'(s)\,ds\wedge\phi_{\theta}^{-1}d_{\theta}\phi_{\theta}
+f(s)(f(s)-1)(\phi_{\theta}^{-1}\,d_{\theta}\phi_{\theta})^2
\right\rb{}P^U
\label{42} \\
(F_{\wnabla})^n&=&nf'(s)(f(s)(f(s)-1))^{n-1}\,ds\wedge
(\phi_{\theta}^{-1}\,d_{\theta}\phi_{\theta})^{2n-1}P^U
\label{43}
\ee
where we have used the fact that $\phi_{\theta}^{-1}d_{\theta}\phi_{\theta}$
maps $\hC_+^U$ to itself, i.e. 
$\lb\phi_{\theta}^{-1}d_{\theta}\phi_{\theta}\,,P^U\rb=0$.
After substituting these in (\ref{39}), the radial parameter in the second 
integral can be integrated out to obtain
\be
Q_{2n}&=&\frac{1}{(2\pi{}i)^n\,n!}\left(\ \int_{B^{2n}}
\Tr\left\lb{}P(dP)^{2n}\right\rb
\;+\;\chi(n)\int_{S^{2n-1}}\Tr\left\lb{}P^U
(\phi_{\theta}^{-1}d_{\theta}\phi_{\theta})^{2n-1}\right\rb
\ \right) \nonumber \\
&&\label{4.12}
\ee
where
\be
\chi(n)=-n\int_0^1f'(s)(f(s)(f(s)-1))^{n-1}\,ds
=(-1)^n\,\frac{n!(n\!-\!1)!}{(2n\!-\!1)!}
\label{4.13}
\ee
To get the second equality in (\ref{4.13}), note that the integral depends
only on the values of $f(s)$ at $s\!=\!0$ and $s\!=\!1\,$; to evaluate it
we can therefore replace $f(s)$ by $s$.

The first term in (\ref{4.12}) is the first term in (\ref{4.1}).
Substituting $P^U=\frac{1}{2}(1-\e^U)$ the second term in
(\ref{4.12}) becomes
\be
\frac{\chi(n)}{2(2\pi{}i)^nn!}\left(\,
\int_{S^{2n-1}}\Tr\left\lb
(\phi_{\theta}^{-1}d_{\theta}\phi_{\theta})^{2n-1}\right\rb
\;-\;\int_{S^{2n-1}}\Tr\left\lb\e^U
(\phi_{\theta}^{-1}d_{\theta}\phi_{\theta})^{2n-1}\right\rb\right)\,.
\label{4.14}
\ee
The second term here is the second term in (\ref{4.1}). The first term 
here can be rewritten as
\be
\frac{\chi(n)}{2(2\pi{}i)^nn!}\int_{S^{2n-1}}\sum_x4\,\tr\left\lb
(\phi_{\theta}(x)^{-1}d_{\theta}\phi_{\theta}(x))^{2n-1}\right\rb
\label{4.15}
\ee
where $\tr$ is the trace for the fundamental representation of SU(N) and
the factor $4$ comes from the trivial trace over spinor indices.
To show that this coincides with the last term in (\ref{4.1}), thereby 
completing the proof of the Theorem, it remains to show that
\be
\frac{\chi(n)}{(2\pi{}i)^nn!}\int_{S^{2n-1}}\tr\left\lb
(\phi_{\theta}(x)^{-1}d_{\theta}\phi_{\theta}(x))^{2n-1}\right\rb
=deg(\phi(x))
\label{4.16}
\ee
To get this, note that 
\be
deg(\phi(x))=\frac{-1}{(2\pi{}i)^nn!}\int_{\wB^{2n}}
\tr\,\hat{F}^n
\label{4.17}
\ee
where $\hat{F}$
is the curvature of the gauge field
$\hat{A}(\theta,s)=f(s)\phi_{\theta}^{-1}(x)\,d_{\theta}
\phi_{\theta}(x)$
on $\wB^{2n}\,$, since this is gauge-equivalent to zero 
at the boundary $S^{2n-1}$ of $\wB^{2n}$ via the 
gauge transformation $\theta\mapsto\phi_{\theta}^{-1}(x)$.
After integrating out the radial parameter in $\wB^{2n}\,$, (\ref{4.17})
reduces to the left-hand side of (\ref{4.16}).\footnote{Note that 
$\hat{F}^n$ is given 
by (\ref{43}) without the $P^U$ and with 
$\phi_{\theta}\to\phi_{\theta}(x)$.}
This completes the proof of Theorem 1. \qed

\section{Classical continuum limit}

In this section
we consider the situation where the $2n$-sphere $\{U^{(\theta,t)}\}$ in
$\U/\G$ is the lattice transcript of a generic $2n$-sphere 
$\{A^{(\theta,t)}\}$ in the orbit space $\A/\G$ of 
smooth continuum SU(N) gauge fields on
the 4-torus. I.e. $A^{(\theta,1)}=\phi_{\theta}\cdot{}A\,$ 
($A\equiv{}A^{(0,1)}$) where $\{\phi_{\theta}:T^4\to
\mbox{SU(N)}\ ,\ \theta\in{}S^{2n-1}\}$ corresponds to a smooth map
$\Phi:S^{2n-1}\times{}T^4\to\mbox{SU(N)}$ via 
$\phi_{\theta}(x)=\Phi(\theta,x)$. In this case it is well-known that the
topological charge of the continuum Dirac index bundle
over the $2n$-sphere in $\A/\G$ equals the degree $deg(\Phi)$ of the map
$\Phi$. (This follows, e.g., from the result in \cite{AS}.)
We announce and outline the argument for the following result, which will
be shown in \cite{prep}:

\vspace{1ex}
\noindent {\em Theorem 2}. (i) The first two terms in the lattice families
index formula (\ref{4.1}) reduce to the topological charge of the continuum
Dirac index bundle over the $2n$-sphere in $\A/\G$ in the classical continuum
limit, i.e.
\be
&&\lim_{a\to0}\;
{\textstyle \frac{1}{(2\pi{}i)^n}}
\left({\textstyle \frac{1}{n!}}\int_{B^{2n}}
\Tr\left\lb{}P(dP)^{2n}\right\rb 
\,-\,{\textstyle \frac{(-1)^n}{2}\frac{(n-1)!}{(2n-1)!}}
\int_{S^{2n-1}}\Tr\left\lb\e^U
(\phi_{\theta}^{-1}d_{\theta}\phi_{\theta})^{2n-1}\right\rb
\ \right) \nonumber \\
&&=\;deg(\Phi) \label{5.1}
\ee

\noindent (ii) When the gauge transformations are constrained to belong to
$\G_0:=\{\phi\in\G\,|\,\phi(x_0)=1\}$, where $x_0$ is an arbitrary
basepoint in $T^4$, the remaining term $2\sum_xdeg(\phi(x))$ in the lattice
families index formula (\ref{4.1}) vanishes. 
\vspace{1ex}

\noindent {\em Corollary}. 
The topological charge of the index bundle of the
Overlap lattice Dirac operator over the lattice transcript of a generic
$2n$-sphere in the continuum orbit space $\A/\G_0$ coincides 
with the continuum topological charge $deg(\Phi)$ when the lattice is 
sufficiently fine.
\vspace{1ex}

Theorem 2 (i) and (ii) together imply that the topological charge of the 
lattice index bundle reduces to that of the continuum index bundle over
the $2n$-sphere in $\A/\G_0$ in the $a\to0$ limit (classical continuum limit).
Since the lattice and continuum topological charges are both integers, 
the Corollary follows. We remark that the gauge field orbit space specified
by the constrained gauge transformations is a smooth manifold, since $\G_0$ 
(unlike $\G$) acts freely on the space of gauge fields. (For this 
reason the gauge transformations were constrained to belong to $\G_0$ 
in the study of the continuum Dirac index bundle in \cite{AS}.)

In the following we outline the proof of Theorem 2 (the details will be 
given in \cite{prep}). We begin by considering the term 
$2\sum_xdeg(\phi(x))$ in the lattice families index formula (\ref{4.1})
in the present case where the $2n$-sphere in the lattice orbit space
is the lattice transcript of the $2n$-sphere in the continuum orbit 
space. I.e. $\phi_{\theta}$ is the lattice transcript of a $S^{2n-1}$-family
of continuum gauge transformations, also denoted $\phi_{\theta}$, and
$U$ is the lattice transcript of $A$.
Since $\phi_{\theta}(x)$ depends smoothly
on $x$, and $T^4$ is connected, the degree $deg(\phi(x))$ of the map
$S^{2n-1}\to\mbox{SU(N)}\,$, $\,\theta\mapsto\phi_{\theta}(x)$ is independent
of $x$. Denoting this by $deg(\phi)$, we therefore have 
$2\sum_xdeg(\phi(x))=2{\cal N}\,deg(\phi)$ where ${\cal N}$=the number of 
lattice sites. Thus this term diverges in the continuum limit if 
$deg(\phi)\,\ne\,0$, but vanishes when $deg(\phi)=0$.
The latter occurs when the gauge transformations belong to $\G_0\,$, since in
this case $\phi_{\theta}(x_0)\!=\!1$ for all $\theta\in{}S^{2n-1}$, hence
$deg(\phi)=deg(\phi(x_0))=0$. This shows Part (ii) of the Theorem.

The proof of Part (i) of the Theorem is based on the following technical 
result, which we announce here and prove in \cite{prep}:
\be
&&\lim_{a\to0}\ \int_{B^{2n}}\Tr\left\lb(dP)^{2n}P\right\rb \nonumber \\
&&\qquad=\frac{1}{(2\pi{}i)^2}\,\int_{B^{2n}\times{}T^4}\tr\left\lb\,
d_tA_{\mu}^{(\theta,t)}(x)\,dx^{\mu}\wedge(\,d_{\theta}A_{\nu}^{(\theta,t)}(x)
\,dx^{\nu})^{n-1}\wedge{}F^{(\theta,t)}(x)\,\right\rb \nonumber \\
&&\label{45}
\ee
and
\be
&&\lim_{a\to0}\ \int_{S^{2n-1}}\Tr\left\lb(\phi_{\theta}^{-1}\,d_{\theta}
\phi_{\theta})^{2n-1}\e^U\right\rb \nonumber \\
&&\qquad=\frac{1}{2(2\pi{}i)^2}\,\int_{S^{2n-1}\times{}T^4}\tr\left\lb\,
(\phi_{\theta}^{-1}(x)\,d_{\theta}\phi_{\theta}(x))^{2n-1}\wedge{}F(x)^2
\,\right\rb 
\label{46}
\ee
where $F^{(\theta,t)}$ and $F$ are the curvatures of $A^{(\theta,t)}$ and
$A\,$ ($=A^{(0,1)}$), respectively. In the $n\!=\!1$ case this has essentially
already been derived in a related context in \cite{L(nonabelian),DA(NPB)}.  
The derivation for general $n$ is essentially a generalisation of the one
given there. Now, substituting these into the left-hand side of (\ref{5.1})
gives
\be
\frac{1}{(2\pi{}i)^{n+2}}\left\lb\ {
\frac{1}{n!}\int_{B^{2n}\times{}T^4}\tr\left\lb\,
d_tA_{\mu}^{(\theta,t)}(x)\,dx^{\mu}\wedge(\,d_{\theta}A_{\nu}^{(\theta,t)}(x)
\,dx^{\nu})^{n-1}\wedge{}F^{(\theta,t)}(x)\,\right\rb
\atop -\ \frac{(-1)^n}{4}\frac{(n-1)!}{(2n-1)!}
\int_{S^{2n-1}\times{}T^4}\tr\left\lb\,
(\phi_{\theta}^{-1}(x)\,d_{\theta}\phi_{\theta}(x))^{2n-1}\wedge{}F(x)^2
\,\right\rb } \right\rb 
\label{47}
\ee
On the other hand, the degree of $\Phi$ can be expressed as
\be
deg(\Phi)=\frac{1}{(2\pi{}i)^{n+2}\,(n\!+\!2)!}\,
\left\lb\ \int_{B^{2n}\times{}T^4}
\tr\,\F^{\,n+2}\;-\;\int_{\wB^{2n}\times{}T^4}\tr\,\wF^{\,n+2}\,\right\rb
\label{48}
\ee
where $\F\,$, $\wF$ are the curvatures of the gauge fields
\be
\A(\theta,t,x)=A_{\mu}^{(\theta,t)}(x)\,dx^{\mu}\ ,\quad
\wA(\theta,s,x)=A_{\mu}(x)\,dx^{\mu}+f(s)\phi_{\theta}^{-1}(x)\,d_{\theta}
\phi_{\theta}(x)
\label{49}
\ee
on $B^{2n}\times{}T^4$ and $\wB^{2n}\times{}T^4\,$, respectively, since these
are related at the common boundary $S^{2n-1}\times{}T^4$ by the 
gauge transformation $\Phi:S^{2n-1}\times{}T^4\to\mbox{SU(N)}$.
From (\ref{49}) we calculate
\be
\F^{\,n+2}&=&(n\!+\!1)(n\!+\!2)\,d_tA_{\mu}^{(\theta,t)}(x)\,dx^{\mu}\wedge
(\,d_{\theta}A_{\nu}^{(\theta,t)}(x)\,dx^{\nu})^{\,n-1}\wedge{}
F^{(\theta,t)}(x) \label{50} \\
\wF^{\,n+2}&=&(n\!+\!1)(n\!+\!2)\frac{n}{2}
f'(s)(f(s)(f(s)-1))^{n-1}F(x)^2\wedge
{}ds\wedge(\phi_{\theta}^{-1}(x)\,d_{\theta}\phi_{\theta}(x))^{\,2n-1}
\nonumber \\
&&\label{51}
\ee
After substituting these in (\ref{48}) and integrating out the radial 
parameter in $\wB^{2n}$ in the second integral, (\ref{47}) is
obtained, thereby establishing (\ref{5.1}). \qed

\section{Relation to gauge anomalies in lattice chiral gauge theory}

In the continuum, the chiral fermion determinant is not really a function
of the gauge field, but rather a section in the U(1) determinant line bundle
associated with the index bundle of the Dirac operator over the space of
gauge fields (see, e.g., \cite{AS,AG} for the details of this and the 
following remarks). To get the determinant as a function of the gauge field, 
a choice of trivialisation of the determinant line bundle must be made.
Different choices of trivialisations correspond to different 
(gauge field-dependent) complex phase choices for the determinant.
By gauge covariance, the determinant line bundle descends to a line bundle
over the orbit space of gauge fields, and the trivialisations which lead to a
gauge-invariant chiral fermion determinant are precisely those which descend 
to trivialisations of the line bundle over the orbit space.
Therefore, the obstructions to getting a gauge-invariant chiral fermion
determinant are precisely the obstructions to trivialising the determinant
line bundle over the orbit space. The primary obstructions are the 
obstructions to trivialising the line bundle over the 2-spheres in the orbit
space (cf. the discussion in the final section of \cite{AG}). 
Since the Chern character of the
determinant line bundle coincides with the first Chern character 
$ch_1(\index\sd)$ of the index bundle (i.e. the degree 2 part of the total 
character $ch(\index\sd)\,$), these obstructions are the 
topological charges of the index bundle over the 2-spheres in the orbit space.
For the generic 2-spheres considered in the preceding section, which 
correspond to maps $\Phi:S^1\times{}T^4\to\mbox{SU(N)}$, the topological 
charge coincides with the winding number obstruction studied in \cite{AG}
(i.e. the winding number of the phase of 
$\det(\sd_+^{\ \phi_{\theta}\cdot{}A})$ as $\theta$ goes around $S^1$).

The situation for the lattice version of the chiral fermion determinant
arising in the overlap formulation \cite{ov} is completely analogous.
It is again a section in a U(1) determinant line bundle \cite{ov,Neu(geom)}
(see also \cite{DA(NPB)} for further discussion).
The line bundle is 
$\Lambda^{max}\hC_+\otimes(\Lambda^{max}\C_-)^*$.\footnote{In fact the 
lattice version of the chiral fermion determinant in the overlap formulation
equals the determinant of $D:\hC_+\to\C_-$ where $D$ is the overlap Dirac
operator \cite{Kiku(overlap),DA(NPB)}.}
But this is precisely the determinant
line bundle associated with the index bundle (\ref{29}) of the Overlap lattice
Dirac operator that we have introduced in \S3.
Thus the primary obstructions to the existence of a gauge-invariant phase 
choice for the overlap chiral fermion determinant are 
the topological charges of the lattice index bundle over the 
2-spheres in the orbit space of lattice gauge fields. For generic 2-spheres,
these coincide with the winding number obstructions for the overlap determinant
studied in \cite{DA(NPB)}. Indeed, the formula for the topological charge 
$Q_2$, given by the $2n\!=\!2$ case our families index formula (\ref{4.1}), 
coincides with the winding number formula Eq. (3.11) of \cite{DA(NPB)}.

When the gauge group is SU(N) with N$\ge$3 there are always nonvanishing
obstructions of the primary type discussed above. This is because of the
existence of topologically nontrivial maps 
$\Phi:S^1\times{}T^4\to\mbox{SU(N)}$ for all N$\ge$3.\footnote{The topological
structure of SU(N) is $S^3\times{}S^5\times\cdots\times{}S^{2N-1}$ modulo
a finite set of equivalence relations. Thus when N$\ge$3 these is always an
$S^5$ factor and topologically nontrivial maps 
$\Phi:S^1\times{}T^4\to\mbox{SU(N)}$ can be constructed from maps
$S^1\times{}T^4\to{}S^5$ with nonvanishing degree.}
Hence the determinant line bundle over the orbit space cannot be trivialised
in these cases and no gauge-invariant phase choice for the chiral fermion 
determinant (with fermion in the fundamental representation) exists.
The same is true in the lattice setting, at least when the lattice is
sufficiently fine, due to the classical continuum limit result of 
\cite{DA(NPB)} (or equivalently, the $2n\!=\!2$ case of Theorem 2 of the 
present paper). 

On the other hand, for gauge group SU(2) there are no topologically nontrivial
maps $\Phi:S^1\times{}T^4\to\mbox{SU(2)}$. Hence the primary obstructions
all vanish in the continuum setting. There might still be nonvanishing primary
obstructions in the lattice setting though; all we can say for certain at
present is that they vanish in the classical continuum limit.
However, there is another obstruction, namely Witten's global gauge
anomaly \cite{Witten(SU(2))}.\footnote{The nonvanishing of Witten's global
anomaly in the continuum in the SU(2) case is related to the fact that the 
space of maps $T^4\to\mbox{SU(2)}$ has two connected components, i.e. there 
are maps (gauge transformations)
which cannot be continuously deformed to the trivial map 
(see \cite{Witten(SU(2))}).} The presence of this obstruction in the
SU(2) theory in the lattice setting has been demonstrated both numerically
\cite{Neu(SU(2)),BC} and analytically 
(in the classical continuum limit) in \cite{Bar}. 
This obstruction also has a natural description in the context
of families index theory for the Dirac operator: 
There is a canonical trivialisation of the U(1) 
determinant line bundle over 1-dimensional balls (i.e. line segments) in
the space of gauge fields; it specifies a 
{\em real} line bundle which descends to a real line bundle over circles
in the orbit space with structure group $\mbox{O(1)}\cong{\bf Z}_2$. (The 
circles come from line segments with gauge equivalent end points.)
The global gauge anomaly is the obstruction to trivialising this bundle.
In the continuum setting this description is implicit
in \cite{Witten(SU(2))}. This viewpoint on the global anomaly, and related 
issues, will be discussed in detail in the lattice setting in 
\cite{DA(global)}.

\section{Concluding remarks}

\noindent (1) We have shown that the index bundle of the Overlap lattice
Dirac operator over the orbit space of SU(N) lattice gauge fields can be 
constructed in a natural way, and have derived a formula (``lattice families 
index theorem'') for its topological charge over generic $2n$-spheres in 
the orbit space. An unanticipated feature is the presence (when 
$2\le{}n\le{}N$) of an integer-valued term which has no continuum analogue
and which generally diverges in the 
classical continuum limit. However, when the gauge transformations are 
constrained to belong to $\G_0=\{\phi\in\G\,|\,\phi(x_0)=1\}$ (as was also 
done in \cite{AS} to make the orbit space a smooth manifold), this term 
vanishes for the $2n$-spheres in $\U/\G_0$ which are lattice transcripts
of $2n$-spheres in the continuum orbit space $\A/\G_0$. The rest of the
formula reduces in the classical continuum limit
to the topological charge of the continuum index bundle over $2n$-sphere
in the continuum orbit space. (This was announced and outlined here; the
key technical part of the argument will be given in \cite{prep}.)
Thus we have seen how topology of the continuum Dirac index bundle over
$\A/\G_0$ can be captured in a finite-dimensional lattice setting.

An implication of this is that $2n$-spheres in $\U/\G_0$
which arise as lattice transcripts of noncontractible $2n$-spheres in 
$\A/\G_0$ (i.e. those with $deg(\Phi)\ne0$) 
are again noncontractible. (For if this were
not the case, the lattice index bundle over the $2n$-sphere would be 
trivialisable and hence have vanishing topological charge, in contradiction 
with the classical continuum limit result, Theorem 2.)
All this provides further evidence that the
orbit space of lattice gauge fields is a good finite-dimensional model
for the orbit space of continuum SU(N) gauge fields on an
even-dimensional torus,\footnote{Earlier evidence for this came from the 
demonstrations that the lattice theory reproduces the global gauge anomaly
and obstructions to the vanishing of local gauge anomalies
\cite{Neu(SU(2)),Neu(geom),BC,Bar,DA(NPB)}.} 
the topology of which is of considerable mathematical
interest and potential physical relevance. We emphasize that,
a priori, it is not at all clear that this should be the case. The situation
is complicated by the fact that the measure-zero subspace of lattice
gauge fields $U$ for which $H^U$ has zero-modes needs to be excised from
$\U$, and it is difficult to say anything at all about what the resulting
topological sectors of $\U$ look like.
Although the lattice orbit space seems to be reproducing the topology of
the continuum one, the situation before modding out by the gauge 
transformations is quite different: In the continuum, the topological
sectors are just affine $\infty$-dimensional vector spaces with no nontrivial
topology; however, in \cite{DA(app)} we will show that the 
topological sectors of $\U$ do have nontrivial topology 
no matter how fine the lattice is:
Using the results of the present paper we will show that they contain 
noncontractible spheres of various dimensions. These are not simply due to 
the presence of noncontractible spheres in SU(N) but directly reflect the
topological structure of $\U$ resulting from excising the $U$'s for which
$H^U$ has zero-modes. (E.g. in the SU(2) case we find that there are 
noncontractible circles.) This is closely connected with
the existence question for $\G_0$ gauge-fixings in the lattice theory, 
which we also discuss in \cite{DA(app)}.

\noindent (2) The topological charge of the continuum Dirac
index bundle over a $2n$-sphere in the orbit space can be expressed as the 
index of a Dirac operator in $2n\!+\!4$ dimensions, or $2n\!+\!2m$ dimensions 
when the dimension of the spacetime is $2m$ (see, e.g., \cite{AG}). 
There is an analogous result in the present lattice setting. It has already 
been given in the $n\!=\!1$ case in \cite{DA(PRL)}; the generalisation to
arbitrary $n$ is straightforward and will be given in \cite{prep}.

\noindent (3) The construction of the lattice index bundle, and the 
derivation of the families index formula (\ref{4.1}), go through for
general ``Overlap-type'' lattice Dirac operators of the form
$D^U=\frac{1}{a}(1+\g5\e^U)$ where $\e^U:\C\to\C$ is any operator
depending smoothly and gauge-covariantly on $U$ and
with the properties $(\e^U)^2=1$ and $(\e^U)^*=\e^U$.
Such $D$ are precisely the solutions to the GW relation (\ref{15a})
with the $\g5$ hermiticity property (\ref{15b}). Besides the Overlap Dirac
operator (\ref{14}), another solution is the lattice Dirac operator
resulting from the perfect action approach of Ref. \cite{H}, although this
is given via recursion relations and no closed form expression is known.
Other solutions have been presented in \cite{Chiu}. However, at present we
can only say with certainty that the classical continuum limit result,
Theorem 2, holds for the case where $D$ is the Overlap Dirac operator.

\noindent (4) From a mathematical viewpoint, an obvious question is how to
generalise the constructions and results of this paper to spacetime
manifolds other than the tori. The problem is to find suitable 
generalisations of the ``naive'' lattice Dirac operator $\Sn^U$ in 
(\ref{7}) and the lattice Laplace operator $\Delta^U$ in (\ref{10});
these can then be fed into the formulae (\ref{9})--(\ref{14}) to get
the discrete Dirac operator $D^U$ and the constructions and derivation of
the families index formula go through as before, cf. Remark 3 above.
(A suitable range for the parameter $m$ would also need to be determined.)
Given a polyhedral cell decomposition of a general spacetime manifold, 
a discrete Laplace operator $\Delta$ is easy to construct (e.g. along the 
lines of \cite{Dodzuik}). 
Constructing the discrete $\Sn$ seems less easy though.
For this one needs to find a way to incorporate the spin structure into
the discrete setting. In the tori case the spin structure is particularly
simple and easily incorporated into the discrete setting with hyper-cubic
cell decomposition, cf. (\ref{7}). How to do this in the general case is
less obvious. This is a problem for future work.

{\em Acknowledgements}. I thank Ting-Wai Chiu and his students for their
kind hospitality at NTU. During this work the author was supported by a 
postdoctoral fellowship from the Australian Research Council and a visiting
position at NTU funded by the Taiwan NSC (grant numbers 
NSC89-2112-M-002-079 and NSC90-2811-M-002-001).

\end{document}